\documentstyle[aas2pp4, tighten, psfig]{article}

\newcommand{\etal}{{et al.}}

\def\gsim{\lower 2pt \hbox{$\, \buildrel {\scriptstyle >}\over {\scriptstyle
\sim}\,$}} 
\def\lsim{\lower 2pt \hbox{$\, \buildrel {\scriptstyle <}\over {\scriptstyle
\sim}\,$}} 

\def\ref#1{{\par\noindent \hangindent=3em\hangafter=1 #1\par}}

\def\incpage{\global\advance\count0 by 1}
\def\ddots{\leaders\hbox to 1em{\hss.\hss}\hfill}

\slugcomment{To appear in the Astrophysical Journal}

\begin{document}

\lefthead{Hamilton, Gotthelf, \& Helfand (1996)}
\righthead{The Detection of Faint X-Ray Flashes}

\title{\bf The Absence of X-ray Flashes from Nearby Galaxies and \\
the Gamma-Ray Burst Distance Scale }

\author{T. T. Hamilton}
\affil{Department of Astronomy}
\affil{California Institute of Technology, Pasadena,~CA 91125}
\affil{Electronic Mail: tth@astro.caltech.edu}

\author{E.\ V.\ Gotthelf\altaffilmark{1}}
\affil{Laboratory for High Energy Astrophysics}
\affil{NASA/Goddard Space Flight Center, Greenbelt,~MD 20771}
\affil{Electronic Mail: gotthelf@gsfc.nasa.gov}

\and

\author{D. J. Helfand}
\affil{Department of Astronomy and Columbia Astrophysics Laboratory}
\affil{Columbia University, 538 West 120th Street, New York,~NY 10027}
\affil{Electronic Mail: djh@carmen.phys.columbia.edu}

\bigskip\bigskip\bigskip

\centerline{To appear in the Astrophysical Journal}

\altaffiltext{1}{\rm Universities Space Research Association}

\clearpage

\begin{abstract}

If typical gamma-ray bursts (GRBs) have X-ray counterparts similar to
those detected by {\sl Ginga\/}, then sensitive focusing X-ray
telescopes will be able to detect GRBs three orders of magnitude
fainter than the detection limit of the Burst and Transient
Source Experiment ({\sl BATSE\/}).  If a substantial portion of
the burst population detected by {\sl BATSE\/} originates in a
Galactic halo at distances greater than or equal to 150 kpc, existing
X-ray telescopes will be able to detect GRBs in external galaxies out
to a distance of at least 4.5 Mpc. As reported in Gotthelf, Hamilton,
\& Helfand (1996) the Imaging Proportional counter (IPC) on board the
{\sl Einstein Observatory} detected 42 transient events with pointlike
spatial characteristics and timescales of less than 10 seconds.  These
events are distributed isotropically on the sky; in particular, they
are not concentrated in the directions of nearby external galaxies.
For halo models of the {\sl BATSE\/} bursts with radii of 150 kpc or
greater, we would expect to see several burst events in observations
pointed towards nearby galaxies.  We see none.  We therefore conclude
that if the {\sl Ginga\/} detections are representative of the
population of GRBs sampled by {\sl BATSE}, GRBs cannot originate in a
Galactic halo population with limiting radii between 150 kpc and 400
kpc. Inasmuch as halos with limiting radii outside of this range have
been excluded by the {\sl BATSE\/} isotropy measurements, our result
indicates that all halo models are excluded.  This result is
independent of whether the flashes we do detect have an astronomical
origin.

\end{abstract}


\section {Introduction}

Although their existence has been recognized for over two decades,
gamma-ray bursts (GRBs) remain enigmatic, their distances and inherent
luminosities uncertain by many orders of magnitude.  In recent years
our understanding has increased enormously as a consequence of the
isotropy and apparent luminosity-function measurements carried out by
the {\sl BATSE\/} instrument onboard the {\sl Compton Gamma-Ray Observatory}
(CGRO) (Meegan \etal\ 1992; Hakkila \etal\ 1994a, 1994b; see Fishman \etal\
1989 for a discussion of the {\sl BATSE\/} experiment). The
preponderance of evidence suggests that GRBs originate at one of two
possible classes of sites, either in an extended Galactic halo or at
cosmological redshifts.  Many workers have developed models in which
the GRBs arise from a halo population at distances of tens to hundreds
of kiloparsecs from the Galactic Center (e.g., Smith \& Lamb 1993;
Podsiadlowski, Rees, \& Ruderman 1995).  In this scenario, the
observed inhomogeneity in the number-size relation is understood as
the result of the finite extent of the halo.  The {\sl BATSE\/}
results have effectively excluded models with limiting halo radii of
less than 150 kpc (Hakkila \etal\ 1994a).  Other workers have proposed
that GRBs originate at cosmological distances.  The inhomogeneity is
then understood as a result of a combination of evolutionary effects
and redshift-induced spectral effects (Paczynski \etal\ 1986; Paczynski
\& Rhoads 1993 and references therein).

In this paper we propose and execute a new test of Galactic halo
models.  We begin with a review of the observed X-ray properties of
GRBs and outline our strategy for using existing X-ray imaging data to
constrain models of the GRB source distribution (\S 2).  We then
define the halo models and construct a catalog of nearby galaxies
whose halos were observed by the {\sl Einstein} IPC.  Section 4
presents our principal result -- the complete absence of bursts from
nearby galaxies -- and uses this to constrain burst distances.  The
final section examines the robustness of our conclusions and
summarizes our results.

\section{X-Rays from GRBs}

If our Galaxy is typical, Galactic halo models predict that external
galaxies will have sources of GRBs similar to those surrounding the
Milky Way, presumably with similar spectral and temporal
characteristics.  GRBs in these external halos would, of course, be
much fainter than those from the halo of our Galaxy.  Because
absorption effects will be insignificant over the distances to nearby
galaxies, however, measurement of the flux of GRBs from the halo of an
external galaxy at a known distance would provide an immediate measure
of the intrinsic burst luminosity and, hence, the distance of Galactic
GRBs.  Similarly, an upper limit on the flux of GRBs in nearby
galaxies provides, in the context of halo models, a lower limit on the
distance of the {\sl BATSE\/} detected bursts.

If we assume that the faintest {\sl BATSE\/} bursts originate at a
distance of 150 kpc, the smallest limiting distance consistent with
the {\sl BATSE\/} isotropy tests, it is clear that bursts from
galaxies well beyond M31 will be fainter than the {\sl BATSE\/} limit
by orders of magnitude. The only instruments that have any chance of
detecting high energy sources at such faint flux levels are focusing
X-ray telescopes such as those carried by {\sl Einstein\/}, {\sl
ROSAT\/} and the {\sl Advanced Satellite for Cosmology and
Astrophysics} ({\sl ASCA}).  Unfortunately, such telescopes are confined to
low energy bands at which {\sl BATSE\/} spectroscopy is non-existent.
We have therefore assumed for the present experiment that typical GRBs
have X-ray spectra similar to the spectra of the GRBs observed by {\sl
Ginga\/} (see below).  The two largest available databases of X-ray
observations are from the {\sl Einstein\/} and ROSAT position
sensitive proportional counters.  To minimize the uncertainty
introduced by the requisite extrapolation from the {\sl Ginga\/} 1.5
to 10 keV band, we analyze {\sl Einstein\/} data in preference to the
somewhat softer photons recorded in the ROSAT database.  Nevertheless,
the typical X-ray flux associated with GRBs is highly uncertain and
remains the greatest source of uncertainty in our experiment.  We
discuss this in more detail in section 4.

{\sl Ginga\/} observed a total of 17 GRBs with a mean flux in the
$1.5-10$ keV band of $\sim 4\%$ that of the gamma-ray flux (Yoshida
\etal\ 1993).  Spectral analysis of the brightest of these bursts
showed a thermal spectrum with a best fit bremsstrahlung temperature
of 1.5 keV (Murakami \etal\ 1990).  While many papers presenting {\sl
Ginga\/} results interpreted them in terms of a blackbody spectral
model, this was motivated by the coincidence that the observed {\sl
Ginga\/} burst flux equaled the flux expected from a blackbody with
the classical neutron star radius of 10 km at a distance of 1 kpc.
The {\it observed\/} {\sl Ginga\/} spectrum is consistent with that of
a 1.5 keV thermal bremsstrahlung continuum.  We adopt this model here,
not because we consider it to have any physical significance, but
because it is a convenient parameterization of the best data available
on GRB X-ray spectra.

The XMON experiment aboard {\sl P78-1} also detected 3-10 keV X-ray
counterparts to GRBs, and found flux ratios similar to those detected
by {\sl Ginga\/} (Laros \etal\ 1984).  Both the {\sl Ginga\/} and XMON
results indicate X-ray fluxes somewhat higher than a naive
extrapolation of the burst power law spectrum observed between 40 and
70 keV (from the composite of all BATSE bursts -- Band \etal\ 1993).
Inasmuch as {\sl Ginga\/} was only sensitive to the hard portion of
this thermal excess, it is possible that the effective temperature is
less than the {\sl Ginga\/} fits with $kT= 1.5$~keV.  In that case,
the X-ray emission in the {\sl Einstein} IPC band would be greater than we
assume here.

No previous experiment has detected absorption of X-rays by material
either local to the burst source or lying along the line of sight.
However, since our working band is softer, significant absorption
could affect {\sl Einstein}-observed bursts.  Because most of the
galaxies we include in our sample lie in regions of the sky in which
Galactic absorption is only $\approx 10^{20} / {\rm cm}^{-2}$ (Stark \etal\
1992), such line-of-sight absorption is not a major issue.  However
absorption local to the burst emitter with a column density in the
range $10^{21} \ {\rm to} \ 10^{23} \ {\rm cm}^{-2}$ could greatly
reduce the source fluence in the IPC band.  Such absorbers would have
to be physically large (and at least several AU away from the burst)
however, in order that the flux from the burst would not fully ionize
the absorber, allowing X-rays through.

The composite {\sl Ginga} GRB X-ray spectrum, folded through the {\sl
Einstein\/} spectral response function using the PIMMS software,
yields an expected count rate in the {\sl Einstein\/} IPC of 540
counts per second for a burst with a flux of $2\times 10^{-7}\ {\rm
erg \ cm}^{-2}$ in the {\sl BATSE\/} band, the limit to which the BATSE
team has calculated a reliable number-size relation.  Since the
background count rate in the IPC is almost always less than one count
per thousand seconds per resolution element, an event of a few counts
in a ten-second interval stands out dramatically and can easily be
detected (Gotthelf, Hamilton, \& Helfand 1996, hereafter Paper I).  We
therefore are sensitive to GRBs out to a distance 30 times greater
than the distance of the faintest {\sl BATSE\/} bursts.

Both {\sl BATSE\/} and our experiment are flux-limited.  The longest
timescale on which {\sl BATSE\/} triggers to record a burst is
1024~ms, and burst durations range from tens of milliseconds to
hundreds of seconds.  We are sensitive to X-ray events primarily on a
timescale of 1 to 10 seconds and have defined the flux limit of our
survey accordingly.  If the X-ray bursts are longer than 10 seconds,
this is a conservative approach, but if the X-ray counterparts of GRBs
frequently had timescales much shorter than 10 s, our flux sensitivity
will be proportionately lower than we have estimated.  We note that
all observed X-ray counterparts of GRBs in fact indicate {\it
longer\/} timescales for the X-ray emission; indeed, some of the burst
we detect have tails extending to ~100 s (see figure 1 of Paper I).
At higher energies, investigators have also noticed a correlation
between softer bursts and longer timescales (Yoshida \etal\ 1989;
Laros \etal\ 1984; Norris \etal\ 1986). While these results are not
conclusive, we consider that, since {\it no\/} observation of a X-ray
counterpart to a GRB has detected such an event with a timescale
shorter than 10 seconds, we are justified in our sensitivity
calculation.

It is possible that events other than GRBs could also produce X-ray
flashes in the IPC (Paper I).  However, it is not necessary that we
understand possible alternative sources of transients in order to test
the definite prediction that halo models make regarding extragalactic
GRB X-ray counterparts.

\section{Halo Models and the Catalog of Galaxies Observed}

\subsection{{Halo Models}}

The principal interest in the absence of X-ray transients from nearby
galaxies is the significance of this non-detection as a test of halo
models of GRB sites.  The crucial question here is the number of
bursts that we would expect to detect if GRBs do originate in the
Galaxy's halo.  This depends of course on the number of bursts in our
Galactic halo and their intrinsic luminosity.  We here make the
extremely conservative assumption that their are {\it no\/} GRBs
fainter than those observed by {\sl BATSE\/}.  We adopt a rate for the
Milky Way of 1500 bursts per year.  We derive this number from the
efficiency calculations of the {\sl BATSE\/} team who estimate that
{\sl BATSE\/} is sensitive to about one quarter of the bursts
occurring at the faintest flux levels (Meegan \etal\  1994), and apply
an additional correction factor of 1.28 to account for our position
off-center in the Galactic halo.

The number of bursts expected to be detected by {\sl Einstein\/} is a
sensitive function of the limiting halo radius in two ways.  If the
{\sl BATSE\/} bursts come from a larger halo, then they are
intrinsically more luminous and their X-ray counterparts could thus be
detected from more distant galaxies.  On the other hand, a more
extended halo means that the surface density of bursts in external
galaxies will be lower.  As a practical matter, given the existence of
a limited set of observations in the {\sl Einstein\/} database, these
two effects work against each other.  If bursts originate in
relatively extended halos, then the number of bursts per unit surface
area per unit time will be less for individual galaxies.  However, a
relatively extended halo implies a relatively high intrinsic burst
luminosity.  Therefore they can be seen at greater distances and more
existing {\sl Einstein\/} fields would be expected to contain bursts.

We have calculated the expected number of bursts for all halo models
consistent with the {\sl BATSE\/} isotropy result.  The adopted lower
limit for the limiting burst distance of 150 kpc follows from the
upper limit on the GRB quadrapole moment with respect to the plane of
the Galaxy, and the upper limit of 400 kpc follows from the upper
limit to the dipole with respect to M31 (Hakkila \etal\ 1994b).  We
calculate the expected surface density of bursts $\rho_s$ at a
distance $r$ from the center of the Galaxy for burst source models in
which

\bigskip 

\centerline{$\rho_s \propto {1 \over 1 + (r/r_c)^\alpha}, \quad\quad\quad\quad r<r_{lim},$}

\bigskip 

\centerline{$ \quad\quad \rho_s=0, \quad\quad\quad\quad\quad\quad r>r_{lim},$}

\bigskip 

{\noindent} where $\alpha \approx 2$, and the population abruptly
cuts off at a radius $r_{lim}$.  This formalism is commonly used in
the analysis of {\sl BATSE\/} data, primarily because it is similar to
models of dark matter distributions that are invoked to explain galaxy
velocity profiles (Fich \& Tremaine 1991; Innanen, Harris, \&
Webbink 1983).  Models in which $\alpha < 2$ can also fit the BATSE
data and may be physically more reasonable; as shown in Hakkila \etal\
(1994a), such models require a larger limiting radius.  We adopt the
model with the conservative assumption that $\alpha = 2$, not because
of any belief in its physical significance, but because the use of
such a model facilitates interpretation of our results in the context
of other GRB studies, especially those interpreting {\sl BATSE\/}
data.

The surface density in such models is not significantly dependent on
the value of $r_c$, the softening parameter in the burst site
distribution.  For all values of $r_c$ substantially less than
$r_{lim}$, the expected projected surface density of sources at the
center of the halo is $\rho_s = 6.9 (D/D_{lim})^2$ per $10^6$ seconds
deg$^{-2}$, where $D_{lim}$ is the maximum distance to which a {\sl
BATSE\/} burst at $r_{lim}$ could be detected.  For a halo limit of
150 kpc implying $D_{lim} = 4.5$~Mpc, we expect one burst every
145,000 seconds in the {\sl Einstein\/} field of view.  A 150~kpc halo
at 4.5~Mpc subtends roughly 10 deg$^2$ as indeed does any halo of size
$r_{lim}$ viewed at the distance corresponding to the limiting
sensitivity.

\subsection{{The Galaxy Catalog}}

A complete list of nearby galaxies with distances from 1 to 12 Mpc and
with $M_v < -16$ was drawn from the {\sl Nearby Galaxies Atlas\/}
(Tully 1989). We have excluded all galaxies that are within 30 arcmin
of a brighter galaxy at the same distance in order to ensure that
satellite galaxies deep within the halo of a larger galaxy are not
counted as independent objects.  We have not applied any weighting by
mass to the galaxies.  In our calculations we have formally assumed
that the typical galaxy we observe has a halo identical to that of the
Milky Way.  The observation times are skewed somewhat towards more
luminous galaxies, which were more likely to be chosen as IPC targets
for reasons unrelated to our search.  This means that assuming all
catalog galaxies to be equal contributors to the burst population is a
conservative assumption with regard to the distribution of bursts.  If
we weighted the galaxies, any plausible scheme would place more weight
on the systems which were in fact most observed.

This does not, however, resolve the question of the overall
normalization of the total galaxy luminosity in our sample.  Gott \&
Turner (1976) estimate that the local density of galaxy optical
luminosity is about 2.75 times the optical luminosity density on large
scales.  Adopting their numbers, we calculate that our assumptions are
equivalent to assuming that the burst/galaxy luminosity ratio for our
sample is approximately 1.6 times the value for the Galaxy.
Specifically, we assume that the total burst-producing material along
the line of sight to our sample galaxies has a ratio to those
galaxies' luminosity 1.6 times as great as the ratio of
burst-producing material within the model radius of the Milky Way to
our Galaxy's luminosity.  This is roughly comparable to assuming that
burst production traces mass and applying standard comparisons of mass
to light ratios for galaxies.  If a substantial fraction of the
intergalactic mass inferred from kinematic studies emits bursts, then
the expected bursts will be correspondingly more numerous.  Trimble
(1987) provides a thorough review of the uncertainties of computing
galactic and intergalactic masses in regions with no visible emission.
The fact that the mass of material far from the luminous regions of
the disk is so uncertain leads us to our simple approach.

We next constructed a database containing all IPC pointings whose
centers lay within 5 degrees of any of the 189 galaxies in our
catalog, thus including both observations that were deliberately
pointed at a nearby galaxy and serendipitous observations in which a
galaxy or part of its putative halo is within the field of view.  A
total of $2.8 \times 10^6$ seconds was accumulated, with most of the time
spent in scheduled observations of well-known nearby galaxies; one
flash was detected.  Since {\sl Einstein\/} detected 18 potentially
astronomical flashes in $1.6 \times 10^7$ seconds this is not statistically
unexpected. This result is not dependent on the arguments used in
Paper I to extract the 18 potentially astronomical events from the
complete list of 42 candidates; none of the 24 likely counter events
fell within the nearby galaxy database.  Table 1 lists the galaxy
positions, distances and the total time that {\sl Einstein\/} spent
observing a putative 400 kpc halo about each galaxy's position.  The
observing times for nearby galaxies are large. However, as explained
above, the expected surface density of bursts is low for nearby
galaxies, and, as a result, most of the contribution to the expected
burst total comes from galaxies near the limiting distance for a
particular halo model.

Since larger halo models imply higher luminosities for the {\sl
BATSE\/} burst sample, we must examine observations of galaxies at
larger distances as the assumed $r_{lim}$ increases.  A 200 kpc radius
halo would produce bursts visible out to 6~Mpc, while a 400~kpc halo
is visible to 12~Mpc and so on.  Similarly, the surface density of
bursts from the halos of galaxies at distances less than that of the
limiting sensitivity is reduced by a factor proportional to the square
of the ratio of the distance to the limiting distance.

\subsection{{Results}}

Table 2 lists the predicted number of {\sl Einstein\/}-detected bursts
for six model halos with different limiting radii. No X-ray flashes
were detected in the halos described by any of these six models.

Column 1 of Table 2 lists the value of $r_{lim}$ in equation 1.
Column 2 lists the limiting distance at which bursts can be detected
by {\sl Einstein\/} if the bursts at the {\sl BATSE\/} flux limit are
at a distance $r_{lim}$. Column 3 lists the total exposure time for
all galaxies with $M_v < -16$ whose halos fall within the field of
view of an {\sl Einstein\/} exposure.  Column 4 lists the adjusted
exposure time.  To compute this quantity, the actual exposure time for
each halo was reduced by the square of the ratio of the halo's
distance to the limiting distance.  Note that if galaxies were
distributed uniformly in space, column 4 would always equal half of
column 3.  Columns 5 and 6 give the number of bursts whose detection
is expected and the probability that no bursts would be detected if
the model applied.  Column 7 lists the probability of no bursts being
detected if the physical extent of the halo were twice the distance at
which {\sl BATSE\/} is able to detect bursts.

For $r_{lim} = 600$ kpc, bursts would be observable from the Virgo
cluster.  We would easily see them, since {\sl Einstein\/} observed in
the direction of the cluster for 4,243,000 seconds, often with
multiple galaxies in the field of view.  Indeed, one burst is seen in
the direction of the Virgo cluster (burst \# 3 in Table 1 of Paper I).
This is consistent with the expected random occurrence rate, and is
inconsistent with the 23 bursts from Virgo we would see if typical
halos had 600~kpc radii.  A halo this large would also produce an
anisotropy in the direction of M31 observable with {\sl BATSE\/}
(Hakkila \etal\ 1994).  Our exclusion of such models is therefore an
independent confirmation of the M31 results.

We have also considered the possibility that the {\sl BATSE\/} does
not sample the entire extent of the Galactic halo.  There is, of
course, no reason why the halo could not extend well beyond {\sl
BATSE}'s sampling distance.  Note that {\sl BATSE}'s non-detection of
a dipole towards M31 excludes halos larger than 400 kpc {\it only} if 
{\sl BATSE\/} is able to detect halos that large.  That is, {\sl BATSE\/}
obviously can not constrain the location of bursts it cannot see.
However, our experiment can test for the existence of halos extended
well beyond the {\sl BATSE\/} limit.  Such halos produce many more
expected bursts in our galaxy sample and can be readily excluded as
seen in Table~2.

We therefore conclude that the GRBs detected by {\sl BATSE\/} are not
associated with X-ray bursts coming from a Galactic halo with a
limiting radius greater than 250 ~kpc and less than 400~kpc, or,
equivalently, from bursts with luminosities between $7 \times 10^{38}\
{\rm ergs\ s}^{-1}$ and $2\times 10^{39} {\rm ergs\ s}^{-1}$ in the
$0.16-3.5$~keV band.  This is the range of halo radii favored by the
analysis of Hakkila \etal\ (1994b).  Although our exclusion of halo
models with limiting radii as small as 150 kpc is only weakly
significant (73\%), this result is much stronger if combined with the
prior result of Hakkila \etal\  (1004b). If the GRBs originate in a 150 kpc
halo, then three independent probabilities must be considered: 1) this
halo radius is at the 90\% confidence contour of Hakkila \etal's
(1994b) result; and 2) our result excludes such a halo with 73\% confidence; 3)
{\sl BATSE\/} must have been fortuitously designed to see most of the way to the
halo's edge but not beyond.  The {\it a priori\/} probability of these
three independent coincidences is approximately 1\%.  That is,
combining our result with that of Hakkila \etal\ (1994b) excludes all
halo models with $\geq 99\%$ confidence.  If we believe that X-ray
counterparts are a common feature of GRBs, this would argue strongly
for a cosmological GRB origin.  The regions of parameter space allowed
by Hakkila (1994b)'s results and ours are illustrated in Figure 1.  As
discussed below this chart uses the conservative and inconsistent
assumption that GRBs are standard candles in both the $\gamma$-ray and
X-ray bands.  Deviation from either of these assumptions results in
the exclusion of halo models with greater confidence.

\section{{Robustness of our Conclusion}}

We consider the uncertainty in the GRB X-ray / $\gamma$-ray flux to be
easily the weakest link in our argument.  Current models for the
production of GRBs in a Galactic halo do not predict a sharp
low-energy cutoff at the {\sl Einstein\/} spectral band.  Indeed a
wide variety of fireball models predict a substantial X-ray excess
above what we have used in our calculations (M\'esz\'aros \& Rees
1993).  However, in the absence of a well-established model for the
GRB production mechanism, the possibility that GRB spectra suddenly
cutoff at the boundary of the {\sl Einstein\/} and {\sl Ginga\/} bands
cannot be excluded.  Unfortunately CGRO does not carry an instrument
capable of measuring the spectra of the faint bursts it detects down
to X-ray wavelengths.  It is likely that in the next few years,
however, new experiments will remedy this lack of knowledge and
establish definitively the X-ray character of the GRBs.

Closely related to the uncertainty in X-ray / $\gamma$-ray flux ratio
is our use of the assumption that GRBs are X-ray standard candles.
This is inconsistent with the assumption of Hakkila \etal\ (1994a)
that $\gamma$-rays from GRBs are standard candles, because the ratio
of X-ray / $\gamma$-ray flux is known to vary widely (Yoshida \etal\
1989; Laros \etal\ 1984).  Moreover, {\sl BATSE\/} reports wide
variation in the spectrum of the $\gamma$-rays it observes (Band
\etal\ 1993).  Given this spectral variability it is highly unlikely
that any experiment would measure exactly a band in which the GRBs
were standard candles.

In the interpretation of both the BATSE and IPC results, non-standard
candles tend to reduce the parameter space available for halo models.
In particular for the IPC result, the assumption of non-standard
candles increases the distance at which some bursts could be detected
for a given halo model. Since the volume of space from which bursts
can be detected with a luminosity $L$ increases as $L^{3/2}$, the
total number of detectable bursts increases.  In the context of our
models this means that bursts four times brighter than average from a,
say, 200 kpc radius model would be detected in the galaxy searches
performed for the 400 kpc radius model.  Inasmuch as the volume
searched in the higher radius models includes many more galaxies,
non-standard candle models are excluded with higher confidence, just
as are the higher radius models.  For the 400 kpc model, a factor of
two excursion above average in luminosity would result in bursts
visible from the Virgo cluster, a result we strongly exclude.
	
Another implicit assumption of our analysis is that the halos of
nearby galaxies resemble that of the Milky Way.  The burster halos we
are searching for are at galactocentric distances far greater than the
visible extent of the galaxy's light.  Consequently, no kinematic
evidence exists relevant to the size or frequency of such halos.  Even
if dark halos were shown to exist about these galaxies, there is no
reason to believe that the distribution of GRB source sites would
trace the mass distribution.  Indeed, halo models that satisfy {\sl
BATSE\/} isotropy constraints show less source concentration toward
the center of the Galaxy than halo models derived from rotation curve
analysis (Hakkila \etal\ 1994).  Because of the fast time scale of
observed GRBs, it is clear they must originate from compact sources.
Most models for a Galactic origin of the GRBs postulate an association
with neutron stars.  Recent observations of neutron star proper
motions suggest that the halo may be populated with high velocity
neutron stars that were created during the course of the star
formation history of the Galaxy; i.e., they are not primordial (Lyne
\& Lorimer 1994).

Support for this hypothesis follows from the recent association of
supernova remnants with soft gamma repeaters (SGRs) (Murakami \etal\
1994).  The SGRs appear to be associated with young, high-velocity
neutron stars (Rothschild, Kulkarni, \& Lingenfelter 1994).  Perhaps
such objects may in time populate an extended halo about any galaxy
with an appropriate history of supernovae.  If this is the case, it is
not completely obvious what types of galaxies would have what types of
halos.  There may be a complex relationship between mass, galaxy type
and halo extent or density.  Knowing little, we have followed a simple
approach.  Because we make the implausible assumption that the {\sl
BATSE\/} detection limit represents an absolute limit on the burst
population -- i.e., that there are no bursts in our Galaxy below the
{\sl BATSE\/} limit -- we consider our estimates to be conservative.
However it is obviously possible that our Galaxy is anomalous with
respect to its GRB source population.

\acknowledgements

T. T. H. acknowledges support from NASA grant NAGW-4110 and wishes to
thank Fiona Harrison, David Hogg, and Stephen Thorsett for useful
discussions. D. J. H. acknowledges support from NASA grant NAS5-32063 and
gwishes to express his gratitude to his local wine merchant whose
case-discount policy has allowed him to avoid bankruptcy in covering
his bets that GRBs were Galactic.

\clearpage

\clearpage

\onecolumn

\voffset=-1.truein
\font\cpr=cmr8
\cpr
\font\apj=cmcsc10 at 9truept
\vsize=10truein

\centerline{\apj Table 1}
\medskip
\centerline{\apj List of Galaxies}
\bigskip\medskip
\hrule height .08em
\vskip 2pt
\hrule
\vskip 1em

\hbox{
\vtop{
\halign{\hfil #\hfil\tabskip1.4em & \hfil# &\hfil# & \hfil#\cr
Right Ascension &\omit\hfil Declination\hfil &\omit\hfil Distance\hfil
&\omit\hfil Time\hfil\cr 
(1950) &\omit\hfil (1950)\hfil &\omit\hfil (Mpc)\hfil
&\omit\hfil (s)\hfil\cr
\noalign{\vskip 1em\hrule\vskip 2em}
\multispan4\hfil  Galaxies at 4--12 Mpc\hfil\cr\noalign{\vskip 5pt}
   00   43    18 &   - 15  52 &     11.6 &       0\cr
   00  49    18 &    47  17 &     11.8 &        0\cr
   01  27    12 &    - 01  30 &     10.6 &    12888\cr
   01  34    00 &    15  32 &      9.7 &     6867\cr
   01  39    42 &    13  43 &     10.8 &        0\cr
   01  40    18 &    13  23 &     11.8 &        0\cr
   01  44    42 &    27  05 &      6.4 &     8339\cr
   01  45    00 &    27  11 &      7.5 &     8339\cr
   01  46    42 &    32  20 &      4.6 &    72619\cr
   01  58    24 &    28  35 &      4.7 &    16197\cr
   02  19    18 &    42  07 &      9.6 &    10364\cr
   02  21    54 &    35  49 &      9.8 &     6335\cr
   02  24    18 &    33  22 &      9.4 &    10474\cr
   02  27    48 &    36  55 &     10.3 &        0\cr
   02  29    18 &    35  17 &     10.1 &     6335\cr
   02  30    18 &    33  17 &     10.1 &     6335\cr
   02  30    36 &    40  19 &     10.2 &     2748\cr
   02  33    24 &    25  13 &     10.7 &     1432\cr
   02  36    06 &    40  40 &     10.7 &     2748\cr
   02  37    18 &    38  51 &     10.5 &     2748\cr
   02  37    42 &    19  05 &     11.2 &     1972\cr
   02  40    12 &    37  08 &      9.1 &     1629\cr
   02  44    48 &    37  20 &     10.0 &     1629\cr
   02  55    24 &   - 54  46 &      5.4 &     5900\cr
   02  56    48 &    25   02 &      6.4 &     9961\cr
   03  08    36 &   - 53  32 &     10.7 &     5900\cr
   03  15    30 &   - 41  19 &      8.6 &     2809\cr
   03  24    18 &   - 52  57 &     11.5 &    10890\cr
   03  30    12 &   - 52  05 &     11.6 &     6436\cr
   03  31    54 &   - 31  22 &     11.6 &      562\cr
   03  37    06 &   - 44  15 &     11.2 &     2016\cr
   03  37    18 &   - 18  51 &      5.0 &    26936\cr
   03  37    30 &   - 31  30 &     11.8 &     1646\cr
   03  40    30 &   - 47  23 &     11.6 &        0\cr
   03  55    54 &   - 46  21 &     11.3 &     2016\cr
   04  01    54 &   - 02  19 &     10.6 &      915\cr
   04  01    54 &   - 43  33 &     10.3 &     2052\cr
   04  02    18 &   - 43  29 &      9.5 &     2052\cr
   04  06    54 &   - 48  01 &     11.0 &        0\cr
   04  38    54 &   - 02  56 &      8.9 &     5975\cr
   04  53    06 &   - 53  27 &      6.0 &    14185\cr
   04  57    54 &   - 26  06 &      7.8 &     1448\cr
   05  02    06 &   - 61  12 &     10.6 &        0\cr
   05  04    30 &   - 32  01 &      7.4 &        0\cr
   05  06    00 &   - 37  35 &     10.8 &        0\cr
   05  08    48 &   - 31  40 &     10.8 &        0\cr
}}
\hskip .5truein
\vtop{
\halign{\hfil #\hfil\tabskip1.4em & \hfil# &\hfil# & \hfil#\cr
Right Ascension &\omit\hfil Declination\hfil &\omit\hfil Distance\hfil
&\omit\hfil Time\hfil\cr 
(1950) &\omit\hfil (1950)\hfil &\omit\hfil (Mpc)\hfil
&\omit\hfil (s)\hfil\cr
\noalign{\vskip 1em\hrule\vskip 2em}
   05  09    36 &    62  31 &      4.5 &    14821\cr
   05  10    06 &   - 33  02 &     10.2 &        0\cr
   05  13    42 &    53  30 &     11.4 &        0\cr
   05  45    12 &   - 34  15 &     10.2 &     9656\cr
   05  33    06 &    03  24 &     10.3 &        0\cr
   06  08    24 &   - 34  06 &      7.9 &     1921\cr
   03  27    00 &    39  31 &      8.6 &        0\cr
   07  06    36 &    44  32 &      8.2 &     2610\cr
   07  32    06 &    65  43 &      4.2 &    82623\cr
   07  35    00 &   - 47  31 &     10.9 &        0\cr
   07  58    12 &    50  54 &     10.1 &        0\cr
   08  09    42 &    46  09 &      9.0 &     8390\cr
   08  10    24 &    45  54 &     10.6 &        0\cr
   08  11    00 &    49  13 &     10.6 &     8390\cr
   08  14    06 &    70  52 &      4.5 &    57834\cr
   08  15    42 &    50  10 &     10.0 &     8390\cr
   08  49    36 &    33  38 &      5.7 &    13479\cr
   08  55    48 &    39  24 &      8.7 &     4249\cr
   09  04    24 &    33  28 &      7.8 &     1319\cr
   09  10    06 &   - 23  58 &      7.1 &     6102\cr
   09  15    42 &   - 22  09 &     10.8 &     6102\cr
   09  18    36 &    51  12 &     12.0 &        0\cr
   09  29    24 &    21  44 &      6.3 &    11298\cr
   09  51    42 &    69  55 &      5.2 &    37258\cr
  10   00    54 &    41  00 &      9.4 &        0\cr
  10   02    42 &    - 07  29 &      6.7 &     4111\cr
  10  15    12 &    41  40 &      8.7 &      614\cr
  10  16    42 &    45  49 &     10.8 &        0\cr
  10  22    24 &    17  25 &      8.1 &     8496\cr
  10  36    24 &    41  56 &     11.5 &        0\cr
  10  40    48 &    25  11 &      6.1 &        0\cr
  10  41    18 &    11  58 &      8.1 &    29026\cr
  10  43    42 &    02  05 &     10.7 &     2576\cr
  10  44    12 &    12  05 &      8.1 &    29026\cr
  10  45    06 &    14  15 &      8.1 &    20724\cr
  10  45    12 &    12  51 &      8.1 &    20724\cr
  10  45    36 &    12  54 &      8.1 &    26872\cr
  10  48    18 &    13  41 &      8.1 &    28961\cr
  10  48    00 &    76  07 &     10.9 &     3767\cr
  10  49    42 &    36  54 &      7.8 &     9728\cr
  10  57    42 &    14  10 &      6.4 &    42074\cr
  10  57    48 &    29  15 &      7.4 &    14465\cr
  11  01    00 &    29  09 &      7.9 &    16049\cr
  11  03    12 &    00  14 &      7.2 &        0\cr
  11  17    36 &    13  17 &      6.6 &    20902\cr
  11  17    42 &    13  53 &      7.7 &    20902\cr
  11  33    00 &    54  47 &      4.3 &    39312\cr
  11  54    06 &    48  36 &      8.3 &     4617\cr
}}}
\bigskip
\hrule
\vfill

\clearpage

\onecolumn

\voffset=-1.truein
\font\cpr=cmr8
\cpr
\font\apj=cmcsc10 at 9truept
\vsize=10truein
\centerline{\apj Table 1 (Continued)}
\medskip
\hrule height .08em
\vskip 2pt
\hrule
\vskip 1em
\hbox{
\vtop{
\halign{\hfil #\hfil\tabskip1.4em & \hfil# &\hfil# & \hfil#\cr
Right Ascension &\omit\hfil Declination\hfil &\omit\hfil Distance\hfil
&\omit\hfil Time\hfil\cr 
(1950) &\omit\hfil (1950)\hfil &\omit\hfil (Mpc)\hfil
&\omit\hfil (s)\hfil\cr
\noalign{\vskip 1em\hrule\vskip 2em}
  11  56    18 &    30  41 &      8.0 &     9593\cr
  12  01    30 &    32  11 &      9.7 &     3595\cr
  12  03    30 &    47  45 &      8.8 &      533\cr
  12  06    42 &    30  12 &      9.7 &    12809\cr
  12  07    30 &    46  44 &      4.1 &    23065\cr
  12  08    00 &    30  41 &      9.7 &     8471\cr
  12  09    48 &    29  28 &      9.7 &    23093\cr
  12  12    36 &    33  29 &      9.7 &     8669\cr
  12  12    42 &    20  56 &      7.9 &     9176\cr
  12  14    54 &    45  54 &      7.5 &    12264\cr
  12  15    06 &    29  53 &      9.7 &    17095\cr
  12  15    24 &    47  41 &      7.3 &     3120\cr
  12  15    36 &    28  27 &      9.7 &    17095\cr
  12  16    30 &    47  35 &      6.8 &     3120\cr
  12  17    24 &    29  53 &      9.7 &    17095\cr
  12  17    36 &    29  34 &      9.7 &    17095\cr
  12  17    48 &    29  35 &      9.7 &    17095\cr
  12  18    12 &    46  35 &      8.0 &     4788\cr
  12  19    54 &    29  29 &      9.7 &    17095\cr
  12  20    06 &    30  10 &      9.7 &    22317\cr
  12  21    36 &    31  48 &      9.7 &    15083\cr
  12  22    06 &    70  37 &     11.1 &    17904\cr
  12  23    18 &    27  50 &      9.7 &    12295\cr
  12  24    00 &    31  30 &      9.7 &    10022\cr
  12  25    48 &    28  54 &      9.7 &    12295\cr
  12  26    12 &    23  06 &      6.2 &    17168\cr
  12  26    24 &    45  09 &      8.1 &     4255\cr
  12  28    12 &    41  58 &      9.3 &     6523\cr
  12  28    18 &    41  55 &      7.8 &     8693\cr
  12  28    54 &    26  03 &      9.7 &     4200\cr
  12  30    00 &    42  59 &      7.5 &     6523\cr
  12  30    12 &    00  23 &      9.8 &    10561\cr
  12  30    24 &    37  54 &      6.2 &    31129\cr
  12  31    18 &    30  34 &      9.7 &     5222\cr
  12  31    42 &    35  48 &      9.8 &     2311\cr
  12  33    30 &    28  14 &      9.7 &        0\cr
  12  33    48 &    26  15 &      9.7 &     2692\cr
  12  36    42 &    00  16 &      9.6 &     4298\cr
  12  39    12 &    41  25 &      7.3 &     4855\cr
  12  39    48 &    32  49 &      6.9 &    15415\cr
  12  41    36 &    32  26 &      7.2 &    10193\cr
  12  46     6 &    51  26 &      8.0 &     1227\cr
  12  48    36 &    41  23 &      4.3 &    13566\cr
  12  54    18 &    21  57 &      4.1 &    18018\cr
  13  00    42 &   - 17  08 &      7.1 &    20254\cr
  13  01    36 &   - 05  17 &      6.4 &    40986\cr
  13  02    30 &   - 49  12 &      5.2 &    24633\cr
  13  10    00 &    44  18 &      6.0 &     6201\cr
  13  13    30 &    42  17 &      7.2 &     6201\cr
}}
\hskip .5truein
\vtop{
\halign{\hfil #\hfil\tabskip1.4em & \hfil# &\hfil# & \hfil#\cr
Right Ascension &\omit\hfil Declination\hfil &\omit\hfil Distance\hfil
&\omit\hfil Time\hfil\cr 
(1950) &\omit\hfil (1950)\hfil &\omit\hfil (Mpc)\hfil
&\omit\hfil (s)\hfil\cr
\noalign{\vskip 1em\hrule\vskip 2em}
  13  16    18 &   - 20  47 &      6.7 &    12865\cr
  13  22    24 &   - 42  45 &      4.9 &    72177\cr
  13  27    42 &    58  40 &      4.8 &    39344\cr
  13  27    48 &    47  27 &      7.7 &        0\cr
  13  34    12 &   - 29  37 &      4.7 &    53825\cr
  14  01    30 &    54  36 &      5.4 &    62760\cr
  14  03    18 &    53  54 &      6.0 &    61442\cr
  14  09    18 &   - 65  06 &      4.2 &    62192\cr
  14  18    12 &    56  57 &      7.0 &    42017\cr
  15  27    12 &    64  55 &     11.2 &    12942\cr
  17  49    54 &    70  10 &      6.1 &   203860\cr
  18  23    30 &   - 67  01 &      8.9 &     2575\cr
  18  44    06 &   - 65  14 &     10.9 &    17959\cr
  20  33    48 &    59  59 &      5.5 &    27662\cr
  20  47    24 &   - 69  24 &      6.7 &    85610\cr
  21  25    36.\rlap{01} & - 52  59 &     10.6 &        0\cr
  21  33    00 &   - 54  47 &     10.4 &        0\cr
  22  01    30 &    43  30 &     10.5 &    49241\cr
  22  18    18 &   - 46  19 &     11.1 &        0\cr
  23  19    42 &    40  34 &      8.6 &     1850\cr
  23  22    18 &    41  04 &      9.3 &     1850\cr
  23  27    36.\rlap{01} & 40  43 &      9.2 &     1850\cr
  23  31    48 &   - 36  22 &      8.4 &     4142\cr
  23  33    36.\rlap{01} &  - 38  12 &      8.2 &     2630\cr
\noalign{\vskip 3pt}
\multispan4\hfil  Galaxies at 1--4 Mpc\hfil\cr\noalign{\vskip 3pt}
   00  44    36 &   - 21  01 &      2.1 &    33505\cr
   00  45    06 &   - 25  34 &      3.0 &    29926\cr
   00  52    30 &   - 37  57 &      1.2 &   116295\cr
   01  06    42 &    35  27 &      2.4 &   111305\cr
   01  32    54 &   - 41  40 &      3.9 &    46641\cr
   03  17    42 &   - 66  41 &      3.7 &     7988\cr
   03  42    00 &    67  56 &      3.9 &     8081\cr
   04  26    00 &    64  45 &      1.6 &    27389\cr
   04  27    06 &    71  48 &      3.0 &    12895\cr
   07  23    36 &    69  18 &      2.9 &   100820\cr
   09  43    12 &    68   8 &      2.1 &   116847\cr
   09  51    30 &    69  18 &      1.4 &   238546\cr
   09  59    24 &    68  59 &      2.1 &   134772\cr
  10  00    48 &   - 25  55 &      1.8 &    12750\cr
  10  24    48 &    68  40 &      2.7 &    80551\cr
  12  13    06 &    36  36 &      3.5 &    94632\cr
  12  14    18 &    69  45 &      2.2 &   113804\cr
  12  15    00 &    38  05 &      3.1 &   107833\cr
  12  23    24 &    33  49 &      3.6 &   128582\cr
  12  25    48 &    44  22 &      3.0 &    85825\cr
  13  19    06 &   - 36  22 &      3.5 &    63770\cr
  13  37    06 &   - 31  24 &      3.2 &    55790\cr
  17  42    12 &   - 64  37 &      3.0 &    25746\cr
  23  55    18 &   - 32  51 &      2.8 &    37713\cr
}}
}
\bigskip
\hrule
\vfill

\onecolumn

\begin{deluxetable}{c c c c c c c}

\tablenum{2}
\tablecaption{Halo Models \label{tab:var-short}}

\tablehead{
\colhead{$r_{lim}$} & \colhead{$D_{lim}$} & \colhead{Total Time} & 
\colhead{Adjusted Time} & \colhead{Expected} & \colhead{Significance} & \colhead{Probability of} \\
\colhead{(kpc)} & \colhead{(Mpc)} & \colhead{(ksec)} & \colhead{(ksec)} & \colhead{Events} & & \colhead{Doubled Radius}
}

\startdata
150 & 4.5   &  391 & 208 & 1.4 & .239   & \hfill 0.027\nl
200 & 6.0   &  638 & 292 & 2.0 & .135  & \hfill 0.004\nl
250 & 7.5   & 1020 & 470 & 3.2 & .041  & \hfill $<0.0001$\nl
300 & 9.0   & 1290 & 606 & 4.2 & .015  & \hfill $<0.0001$\nl
350 & 10.5  & 1590 & 829 & 5.7 & .003  & \hfill $<0.0001$\nl
400 & 12.0  & 1890 & 915 & 6.3 & .002 & \hfill $<0.0001$\nl
\enddata

\end{deluxetable}

\vfill
\begin{figure} 

\psfig{figure=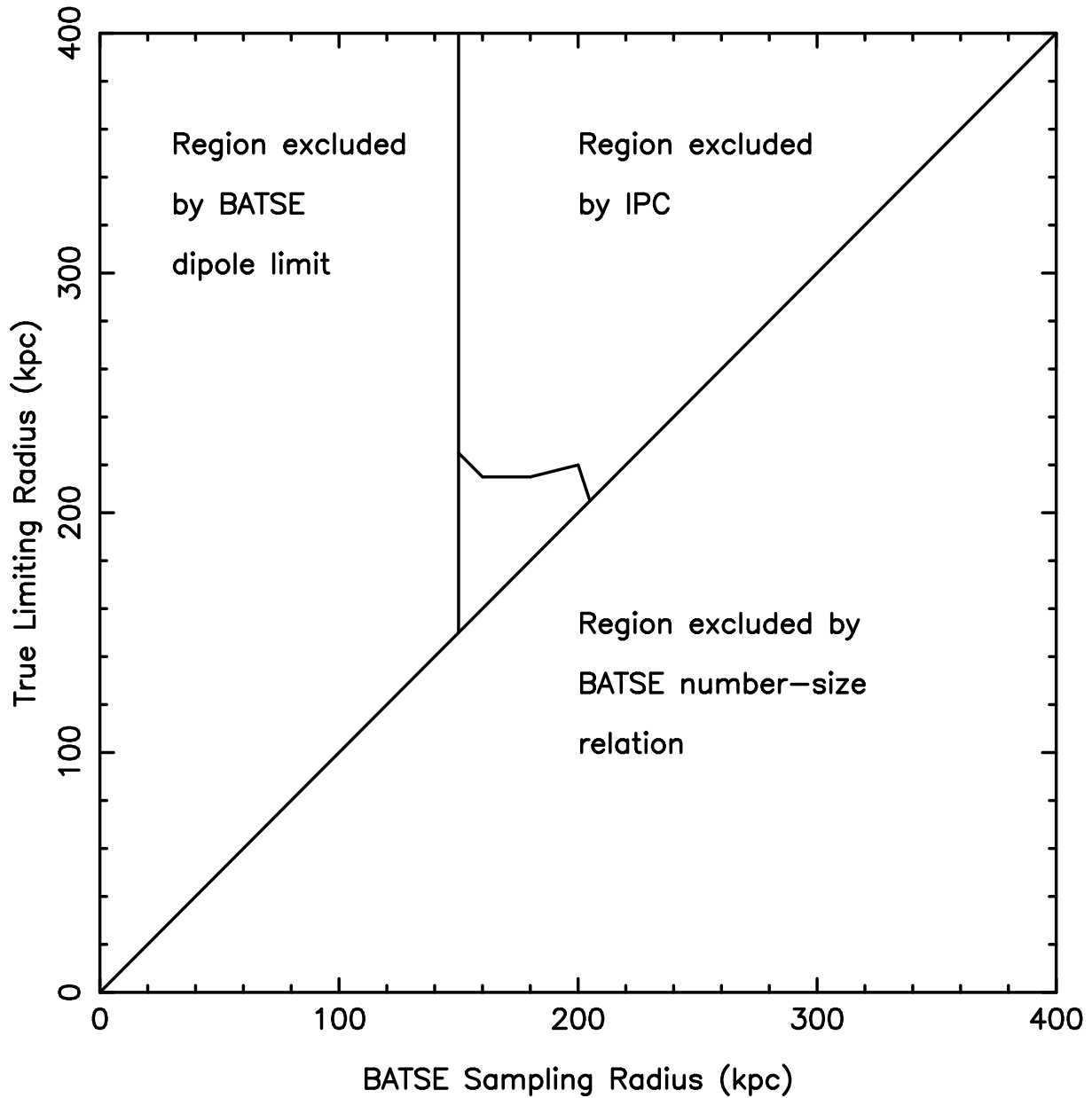,height=18.0truecm,angle=270.0,clip=}

\caption{The region of parameter space of possible halo models excluded with $>
90\%$ confidence by various experiment is plotted. The abscissa is the
distance of the faintest bursts detected by {\sl BATSE\/} and the ordinate is
the distance of the faintest bursts which exist. We have assumed
standard candles and a continuation of the Log $N$ -- Log $S$ below
the {\sl BATSE\/} limit with the same slope. We allow the inner radius
of the distribution to assume any value. For {\sl BATSE\/} models in
the area not excluded by any one experiment, that radius is about 20
kpc. The irregular shape of the IPC contour is a result of the
finite number of nearby galaxies.}
\end{figure}
\vfill
\end{document}